\documentclass[12pt,preprint]{aastex}
\begin{document}

\title{How Do Uncertainties in the Surface Chemical Composition of the Sun Affect the Predicted Solar
  Neutrino Fluxes?}
\author{John N. Bahcall and Aldo  M. Serenelli}
 \affil{Institute for Advanced Study, Einstein Drive,
Princeton, NJ
  08540}

\begin{abstract}
We show that uncertainties in the values of  the surface heavy
element abundances of the Sun are the largest source of the
theoretical uncertainty in calculating the p-p, pep,  $^8$B,
$^{13}$N, $^{15}$O, and $^{17}$F solar neutrino fluxes.  Our
results focus attention on the necessity for improving the
measurement of heavy element abundances while at the same time
reducing the estimated uncertainties in the predicted solar
neutrino fluxes due to abundance errors. We evaluate for the first
time the sensitivity (partial derivative) of each solar neutrino
flux with respect to the surface abundance of each element.  We
then calculate the uncertainties in each neutrino flux using the
preferred `conservative' (based upon changes of measured values
with time) and `optimistic' (current values) estimates for the
uncertainties in the element abundances. The total conservative
(optimistic) composition uncertainty in the predicted $^8$B
neutrino flux is 11.6\% (5.0\%) when sensitivities to individual
element abundances are used. The traditional method that lumps all
abundances into a single quantity (total heavy element to hydrogen
ratio, $Z/X$) yields a larger uncertainty, 20\%. The uncertainties
in the carbon, oxygen, neon, silicon, sulphur, and iron abundances
all make significant contributions to the uncertainties in
calculating solar neutrino fluxes; the uncertainties of different
elements are most important for different neutrino fluxes. The
uncertainty in the iron abundance is the largest source of the
estimated composition uncertainties of the important $^7$Be and
$^8$B solar neutrinos. Carbon is the largest contributor to the
uncertainty in the calculation of the p-p, $^{13}$N, and $^{15}$O
neutrino fluxes. However, for all neutrino fluxes, several
elements contribute comparable amounts to the total composition
uncertainty.
\end{abstract}


\section{INTRODUCTION}
\label{sec:intro}

What is the role of the chemical composition of the surface of the
Sun in the calculation of the solar neutrino fluxes?

In 1966, almost four decades ago, the role of the chemical
composition was described as follows: ``The primordial (or
surface) composition assumed in computing the solar models
represents the largest recognized uncertainty in the predicted
capture rate;...'' (Bahcall 1966)\nocite{bahcall66}. Recently, in
2004, the role of the chemical composition  was summarized as
follows: ``The surface chemical composition of the Sun is the most
problematic and important source of uncertainties.'' (Bahcall \&
Pinsonneault 2004)\nocite{BP04}.

Some things change very slowly.

\subsection{Previous calculations using total heavy element
abundance \boldmath${Z}$}
\label{subsec:previousZ}

 From the very beginning of solar neutrino calculations, the
chemical composition of the Sun has been a principal source of
uncertainty (see Sears 1964)\nocite{1964}. The first systematic
investigations of the effect of composition uncertainties on the
important $^8$B solar neutrino flux concluded that the calculated
neutrino flux was uncertain by a factor of two due to
uncertainties in the chemical composition (Bahcall et
al. 1967\nocite{bahcall67}; Bahcall et al. 1968)\nocite{bahcall2shaviv68}.

In the early days of solar neutrino calculations, the uncertainty in
the neutrino fluxes due to the chemical composition was evaluated by
calculating the sensitivity of the different fluxes to different
choices of the total heavy element abundance (Sears 1964\nocite{1964}; Bahcall
1964;\nocite{bahcall64} Bahcall et al. 1967\nocite{bahcall67}; Bahcall
et al. 1968\nocite{bahcall2shaviv68}; Abraham \& Iben
1971)\nocite{abrahamiben1971}. More refined calculations were
eventually carried out using the logarithmic partial derivatives of
the neutrino fluxes, $\phi_i$, with respect to the total heavy element
abundance by mass, $Z$, i.e., $\partial \ln \phi_i /\partial \ln Z$
(Bahcall \& Ulrich 1971)\nocite{bahcallulrich71}.

The abundances of heavy elements in the solar photosphere are
determined as ratios, the ratio of an individual heavy element
abundance to the abundance of hydrogen, $X$. The abundances are
linked by the relation $X + Y + Z = 1$, where $Y$ is the surface
abundance of helium. The fact that the abundances are measured as
ratios was first taken into account in the paper by Bahcall et al.
(1982),\nocite{bahcalletal82} who calculated partial derivatives
$\partial \ln \phi_i /\partial \ln (Z/X)$. In the intervening two
decades, there have been many evaluations of the uncertainty of
the solar neutrino fluxes caused by the solar composition.
However, these evaluations all used variations with respect  to
the total heavy element abundance, $Z$, or with respect to $Z/X$,
the total heavy element abundance divided by the hydrogen
abundance.

There has not been a previous systematic investigation of the
uncertainty of solar neutrino fluxes due to individual heavy
element abundances. All of the recent papers with which we are
familiar estimate the uncertainty due to the solar composition by
considering variations in the total heavy element abundance, $Z$,
or in $Z/X$ (see, e.g., Schlattl \& Weiss
1999\nocite{SchlattlWeiss1999}; Bahcall et al.
2001\nocite{BP00}; Watanabe \& Shibahashi
2001\nocite{watanabe2001}; Fiorentini \& Ricci
2002;\nocite{fiorentiniricci02} Couvidat et al. 2003\nocite{couvidat03}; Boothroyd \& Sackmann
2003\nocite{boothroydsackmann03}; Bahcall \& Pinsonneault 2004\nocite{BP04};
Young \& Arnett 2005)\nocite{youngarnett}.

Some things change very slowly.

\subsection{Necessity of evaluating uncertainties due to
individual elements}
\label{subsec:individualnecessary}

It has long been recognized that changes in the abundances of some
heavy elements, e.g., iron or silicon, affect the calculated solar
neutrino fluxes more than do the abundances of other elements (see
discussion in \S~IV.D of Bahcall et al.
1982)\nocite{bahcalletal82}. This is largely because the heavier
elements are highly ionized only in the solar core, where they
affect directly the calculated radiative opacity and indirectly
the solar neutrino fluxes.  The lighter, volatile elements affect
the radiative opacity most dramatically in the region somewhat
below the convective zone (temperatures somewhat above $2\times
10^6$ K). Oxygen is a principal contributor to the radiative
opacity just below the convective zone (Turcotte \&
Christensen-Dalsgaard 1998)\nocite{turcottedalsgaard98}.

In recent years, determinations of the solar abundances of heavy
elements have become more refined and detailed (Grevesse \& Sauval
1998\nocite{oldcomp}, 2000; Lodders 2003\nocite{lodders03}) and
especially (Asplund et al.~2000;\nocite{asplundetal00} Asplund
2000;\nocite{asplund00} Allende Prieto et al. 2001,
2002;\nocite{allende01}\nocite{allende02} Asplund et
al.~2004\nocite{asplund04}; Asplund et al.
2005)\nocite{asplundgrevessesauval2005}. These recent
determinations yield significantly lower values than were
previously adopted (e.g., by Grevesse \& Sauval
1998)\nocite{oldcomp} for the abundances of the volatile heavy
elements: C, N, O, Ne, and Ar. However, these recent abundance
determinations lead to solar models that disagree with
helioseismological measurements (Bahcall \& Pinsonneault 2004;
Basu \& Antia 2004)\nocite{basu04}.  For example, the calculated
depth of the convective zone differs by about $15\sigma$ from the
measured value and the calculated surface helium abundance differs
from the measured value by about $7\sigma$ (see equations 1 and 2
of Bahcall, Serenelli \& Basu 2005). Detailed and refined
recalculations of the radiative opacity by the Opacity Project
collaboration disfavor (Seaton \& Badnell 2004\nocite{seaton};
Badnell et al. 2004\nocite{badnelletal2004}) the suggestion (Basu
\& Antia 2004\nocite{basu04}; Bahcall et al. 2004b\nocite{BSP04};
Bahcall et al. 2005)\nocite{BBPS05} that the origin of the
discrepancy might be the adopted opacities rather than the adopted
heavy element abundances.

The discrepancies between helioseismological measurements and the
predictions made using recent determinations of heavy elements make it
especially important that the effect of individual element
uncertainties be evaluated. The discrepancies occur in the temperature
region below the solar convective zone, $2\times 10^6$~K to $4.5 \times
10^6$~K (Bahcall et al. 2005)\nocite{BBPS05}. In this temperature domain, the volatile heavy
elements, C, N, O, Ne, and Ar, are partially ionized and their
abundances significantly affect the radiative opacities. We need to
separate out the effects of the volatile element abundances that
contribute to the helioseismological discrepancies from the effects of
abundances, e.g., Si and Fe, that are most important in the solar
core.

\subsection{What do we do in this paper?}
\label{subsec:planthispaper}

We derive for the first time in this paper individual
uncertainties in each neutrino flux due to each of the important
heavy elements in the solar composition. We then estimate
'conservative' uncertainties by assuming that the differences
between the most recent abundance determinations and the previous
abundance determination represent $1\sigma$ uncertainties. We also
use more 'optimistic' uncertainties taken from the most recent
review of abundance determinations (Asplund et al.
2005)\nocite{asplundgrevessesauval2005}.

We then combine the effects of all composition uncertainties to
determine the net effect of composition uncertainties on each
solar neutrino flux and on the rate of each radiochemical solar
neutrino experiment. Finally, we combine the effects of all known
sources of uncertainties, including composition uncertainties, on
each neutrino flux and experimental radiochemical rate. We
identify the heavy elements that most strongly affect the
predicted neutrino fluxes and we identify for which solar neutrino
fluxes composition uncertainties are most important.

Abundance determinations for the Sun change frequently as improved
techniques, new atomic data, and more observations become
available. Therefore, we describe in this paper the steps
necessary to make detailed and reliable estimates of the
uncertainties in the solar neutrino fluxes for a given set of
abundances and their uncertainties. We also make available at
http://www.sns.ias.edu/$\sim$jnb computer code and numerical data
that can facilitate future investigations when new abundance
determinations are published.

We discuss the role of correlations between the uncertainties of
different elements. This is an aspect of abundance discussions
that is not treated explicitly in any of the papers with which we
are familiar. However, we will show that for future precision
evaluations of the effects of abundance determinations on neutrino
fluxes we must know the correlations, if any, between the quoted
abundance determinations and their uncertainties.

We use for the calculations in this paper the recently computed
standard solar models BP04 and BP04+ (Bahcall \& Pinsonneault
2004)\nocite{BP04}, which are described below. However, we also verify
that small changes such as occur between different recent
redeterminations of the solar abundances cause only negligible changes
(typically $0.1$\%) in the estimated uncertainties of the solar
neutrino fluxes. Table~1 of Bahcall et al. (2004b)\nocite{BSP04} lists the specific element abundances adopted in
computing each of the solar models BP04 and BP04+.

The plan of the paper is as follows. In \S~\ref{sec:logarithmic},
we present and discuss the logarithmic partial derivatives of each
solar neutrino flux with respect to each of the ratios (major
heavy element abundance)/(hydrogen abundance). The results are
given for two separate solar models, BP04 (which uses the Grevesse
 \& Sauval 1988 composition) and BP04+ (which uses more recent
determinations of the abundances of the volatile elements). We
also present the partial derivatives with respect to the heavy
element to hydrogen ratio, $Z/X$. The robustness of the partial
derivatives with respect to $Z/X$ is made evident from the very
small change in their numerical values over more than two decades,
although the solar models have been greatly refined.

We present in \S~\ref{sec:abundanceuncertainties} `conservative'
and `optimistic' estimates for the current uncertainties in the
heavy element abundances. We adopt as our preferred choice the
conservative uncertainty estimates. We compute in
\S~\ref{sec:individual} the uncertainties in individual neutrino
fluxes due to the uncertainties in each heavy element abundance.
We combine for each neutrino flux in
\S~\ref{sec:allcompositionuncertainties} the uncertainties from
all abundance uncertainties; these results are summarized in
Table~\ref{tab:totalcompositionuncertainties}. Our bottom line is
given in Table~\ref{tab:totaluncertainties} of
\S~\ref{sec:alluncertainties}, where we present the total
uncertainties in the neutrino fluxes for different methods of
calculation.

We summarize and discuss our main results in \S~\ref{sec:summary}.

We recommend that all readers start by perusing
\S~\ref{sec:summary}, our summary and discussion section. For many
readers, \S~\ref{sec:summary} contains all they need to know about
the subject.

\section{LOGARITHMIC DERIVATIVES}
\label{sec:logarithmic}

We define in \S~\ref{subsec:defnpartials} the logarithmic partial
derivatives of each of the neutrino fluxes with respect to each of
the element abundances. We present in
\S~\ref{subsec:individualpartials} newly calculated partial
derivatives that were obtained using the recently-derived BP04 and
BP04+ solar models (Bahcall \& Pinsonneault 2004)\nocite{BP04}. We also
calculate in \S~\ref{subsec:zoverxpartials} new values for the
logarithmic partial derivatives with respect to the total heavy
element to hydrogen ratio, $Z/X$. We compare these
newly-calculated values with partial derivatives that were
obtained using 1982 and 1988 solar models. Finally, we use the
results of the previous subsection to explain, in
\S~\ref{subsec:zoverxmisleading},
 the reason why estimates of the
composition uncertainties based upon historical changes in $Z/X$
have led to overestimates of the composition uncertainties in the
calculated solar neutrino fluxes.

\subsection{Definition of partial derivatives}
\label{subsec:defnpartials}

The sensitivity of the neutrino fluxes, $\phi_i$, to the input
parameters,  $\beta_j$, can be expressed to high accuracy in terms
of the logarithmic partial derivatives, $\alpha_{ij}$ (see Bahcall
\& Ulrich 1988\nocite{bahcallulrich1988}; Bahcall 1989\nocite{book}). The logarithmic derivatives are
defined by the equation
\begin{equation}
\alpha_{ij} ~=~ \frac{\partial \ln \phi_i}{\partial \ln \beta_j}\,
.
\label{eq:definitionalphaij}
\end{equation}
In this paper, we are primarily concerned with the uncertainties
in the calculated solar neutrino fluxes that result from
uncertainties in the solar heavy element abundances. Thus we
concentrate on partial derivatives in which the $\beta_j$ are the
mass fractions of different heavy elements relative to the
hydrogen mass fraction. Thus,
\begin{equation}
\beta_j ~=~ \frac{(\rm mass ~fraction ~of ~element ~j)}{(\rm mass~
fraction~ of~ hydrogen)} \, .
 \label{eq:definitionbeta}
\end{equation}
It is conventional to denote the mass fraction of hydrogen by $X$.
\begin{table}[!t]
\caption{\baselineskip=16pt Partial derivatives of neutrino fluxes
with respect to composition fractions. The entries in the table
are the logarithmic partial derivatives, $\alpha_{ij}$ of the
solar neutrino fluxes, $\phi_i$, with respect to the fractional
abundances of the heavy elements, $\beta_j$ (see
eq.~[\ref{eq:definitionalphaij}] and
eq.~[\ref{eq:definitionbeta}]). The partial derivatives were
computed using the solar model BP04 (Bahcall \& Pinsonneault
2004). The derivatives given here are available in digital form
 at http://www.sns.ias.edu/$\sim$jnb under the menu items Solar Neutrinos/\hbox{software and data}. \label{tab:alpha_ij} }
\begin{tabular}{lccccccccc}
\noalign{\smallskip} \hline\hline
\noalign{\smallskip}
Source&C&N&O&Ne&Mg&Si&S&Ar&Fe\\
\noalign{\smallskip}
\hline
pp&$-$0.014&$-$0.003&$-$0.006&$-$0.005&$-$0.005&$-$0.011&$-$0.008&$-$0.002&$-$0.023\\
pep&$-$0.025&$-$0.006&$-$0.011&$-$0.005&$-$0.005&$-$0.014&$-$0.017&$-$0.006&$-$0.065\\
hep&$-$0.015&$-$0.004&$-$0.023&$-$0.017&$-$0.018&$-$0.037&$-$0.028&$-$0.007&$-$0.069\\
$^7$Be&$-$0.002&0.002&0.052&0.049&0.051&0.104&0.074&0.018&0.209\\
$^8$B&0.030&0.011&0.121&0.096&0.096&0.194&0.137&0.034&0.515\\
$^{13}$N&0.845&0.181&0.079&0.057&0.060&0.128&0.094&0.024&0.342\\
$^{15}$O&0.826&0.209&0.093&0.068&0.070&0.150&0.109&0.028&0.401\\
$^{17}$F&0.033&0.010&1.102&0.076&0.078&0.164&0.120&0.031&0.444\\
\noalign{\smallskip} \hline
\end{tabular}
\end{table}

We note that the partial derivatives give rise to the power law
dependences of neutrino fluxes upon model parameters that are
widely used in the literature.  We have
\begin{equation}
\phi_i~=~ \phi_i(0) \left[
\frac{\beta_j}{\beta_j(0)}\right]^{\alpha_{ij}} \, .
\label{eq:powerlaws}
\end{equation}

In practice, the partial derivatives are computed by first
evolving a standard solar model with a specific set of input data.
The standard solar model is used to predict the best-estimate set
of neutrino fluxes, $\phi_i(0)$. Then at least one additional
solar model is evolved in which one parameter, $\beta_i$, is
changed from its standard value. If only these two solar models
are available, then the logarithmic partial derivative
$\alpha_{ij}$ can be estimated from the following equation:
\begin{equation}
\alpha_{ij} ~\simeq~
\frac{\ln\left[\phi_i/\phi_i(0)\right]}{\ln\left[\beta_j/\beta_j(0)\right]}\,
. \label{eq:approximatealphaij}
\end{equation}

 For this paper, we have used five solar models, including the
standard model, in evaluating each of the derivatives
$\alpha_{ij}$. Except for the case of argon, we have  compared
solar models in which the abundance fractions differed from the
standard abundance fractions by -0.10 dex, -0.05 dex, 0.0 dex,
+0.05 dex, and +0.10 dex. Argon is much less abundant than the
other elements we consider;  for this case only, we compared
models with abundance fractions that differed by -0.20 dex, -0.10
dex, 0.0 dex, +0.10 dex, and +0.20 dex from the standard
abundance.

For each modified composition, i.e. for each +-0.05, +-0.10 change
in each of the elements (and +-0.10, +-0.20 for argon), we have
calibrated the corresponding solar model to the present solar
radius and luminosity and to the corresponding Z/X value (the
latter value being different for each modified composition).
Opacity tables were recalculated for each different mixture.

\subsection{The calculated individual partial derivatives}
\label{subsec:individualpartials}

\begin{table}[!t]
\caption{\baselineskip=16pt Same as Table~\ref{tab:alpha_ij} but for a
  solar model with BP04+ (Bahcall \& Pinsonneault 2004) composition.
  The volatile elements C, N, O, Ne, and Ar in this model have the lower
  abundances determined recently (Asplund et
  al.~2000; Asplund 2000;
  Allende Prieto et al. 2001, 2002; Asplund et
  al.~2004) rather than the previously standard
  Grevesse \& Sauval (1998) abundances assumed in
  computing Table~\ref{tab:alpha_ij}.The derivatives given here are available in digital form
 at http://www.sns.ias.edu/$\sim$jnb under the menu items Solar Neutrinos/\hbox{software and data}. \label{tab:alpha_ij_bp04p} }
\begin{tabular}{lccccccccc}
\noalign{\smallskip} \hline\hline
\noalign{\smallskip}
Source&C&N&O&Ne&Mg&Si&S&Ar&Fe\\
\noalign{\smallskip}
\hline
pp & $-$0.010 & $-$0.003 & $-$0.005 & $-$0.003 & $-$0.005 & $-$0.010 &
$-$0.007 & $-$0.001 & $-$0.022 \\
pep & $-$0.018 & $-$0.004 & $-$0.008 & $-$0.002 & $-$0.003 & $-$0.015 &
$-$0.015 & $-$0.003 & $-$0.062 \\
hep & $-$0.012 & $-$0.003 & $-$0.018 & $-$0.011 & $-$0.019 & $-$0.039 &
$-$0.029 & $-$0.005 & $-$0.072 \\
$^7$Be & 0.005 & 0.002 & 0.046 & 0.033 & 0.057 & 0.115 & 0.080 & 0.012 &
0.230 \\
$^8$B & 0.035 & 0.009 & 0.099 & 0.064 & 0.107 & 0.212 & 0.150 & 0.023 & 0.553 \\
$^{13}$N & 0.846 & 0.180 & 0.055 & 0.036 & 0.065 & 0.139 & 0.102 & 0.015 &
0.355 \\
$^{15}$O & 0.824 & 0.211 & 0.068 & 0.043 & 0.077 & 0.163 & 0.119 & 0.018 &
0.423 \\
$^{17}$F & 0.035 & 0.008 & 1.074 & 0.048 & 0.086 & 0.180 & 0.130 & 0.020 &
0.469 \\
\noalign{\smallskip}
\hline
\end{tabular}
\end{table}

Table~\ref{tab:alpha_ij} presents the partial derivatives computed
using the standard solar model BP04 (Bahcall \& Pinsonneault
2004)\nocite{BP04} that assumes the Grevesse \& Sauval
(1998)\nocite{oldcomp} solar abundances.
Table~\ref{tab:alpha_ij_bp04p} presents the partial derivatives
computed using the solar model BP04+ (Bahcall \& Pinsonneault
2004)\nocite{BP04} that uses  recent determinations for the
volatile elements C, N, O, Ne, and Ar rather than the Grevesse \&
Sauval (1998) values for these abundances.

The derivatives of the important $^7$Be, $^8$B, and pp solar
neutrino fluxes with respect to heavy elements like iron are much
larger than the derivatives with respect to the volatile elements
like C, N, and O. The heavier elements are ionized at the higher
temperatures characteristic of the solar interior, where the
neutrinos are formed, while the volatile elements can be ionized
at the lower temperatures that are characteristic of the region
below the solar convective zone. Thus, for a given fractional
uncertainty, the iron, silicon, and sulphur abundances more
strongly affect the calculated neutrino fluxes while the C, N, and
O abundances more strongly affect the comparison with
helioseismology.

\subsection{Partial derivatives with respect to total \boldmath${Z/X}$}
\label{subsec:zoverxpartials}

Table~\ref{tab:alpha_iZoverX} presents the partial derivatives
with respect to the total heavy element to hydrogen ratio, $Z/X$.
The $Z/X$ derivatives are given here for completeness and for
comparison with previous results. The partial derivatives in
Table~\ref{tab:alpha_iZoverX} are defined by the relation
\begin{equation}
\alpha_{i}(Z/X) ~=~ \frac{\partial \ln \phi_i}{\partial \ln
(Z/X)}\, . \label{eq:definitionalphaiZoverX}
\end{equation}
When computing the partial derivatives in
equation~(\ref{eq:definitionalphaiZoverX}), we change $Z/X$ by
multiplying all elements by the same factor, i.e., we increase
(or decrease) the abundance of all heavy elements by the same
fractional amount.

 We present values for the partial derivatives
that are computed not only with the recent BP04 and BP04+ solar
models (columns [2] and [3] of Table~\ref{tab:alpha_iZoverX},
respectively), but also the values that were computed and
published using the 1988 standard solar model (Bahcall \& Ulrich
1988,\nocite{bahcallulrich1988} column [4]) and the 1982 standard
solar model (Bahcall et al. 1982, column [5]).  We note that the
1982 and 1988 models did not include element diffusion, used older
(Los Alamos) opacities and nuclear reaction rates.

The remarkable robustness of the logarithmic partial derivatives
can be seen by eye by comparing the different columns of
Table~\ref{tab:alpha_iZoverX}.  The rms fractional difference
between the partial derivatives ${\partial \ln \phi_i}/{\partial
\ln (Z/X)}$ computed by Bahcall \& Ulrich (1988) and the values
computed using the BP04 solar model is 1.9\%.  Even if we go all
the way back to the first computation, by Bahcall et al.
(1982)\nocite{bahcalletal82}, the rms fractional difference
between the 1982 and the 2004 values (columns [4] and [2],
respectively) is only 10.6\%. The robustness of the partial
derivatives is due in part to the fact that all of the models are
calibrated to the relatively accurately known solar luminosity and
age.

\begin{table}[!t]
\caption{\baselineskip=16pt $Z/X$: Partial derivatives of neutrino
fluxes with respect to the total heavy element to hydrogen ratio.
The entries in the table are the logarithmic partial derivatives,
$\alpha_{i}(Z/X)$ of the solar neutrino fluxes, $\phi_i$, with
respect to the the heavy element to hydrogen ratio $Z/X$ (see
eq.~[\ref{eq:definitionalphaiZoverX}]). The partial derivatives were
computed using the BP04 solar model (column [2]), the BP04+ solar
model (column [3]), the Bahcall \& Ulrich (1988) standard model
(Table~XV of BU88, column [4]), and the Bahcall et al. (1982)
standard model (Table~XI of Bahcall et al. 1982, column [5]).
\label{tab:alpha_iZoverX} }
\begin{center}
\begin{tabular}{lrrrc}
\noalign{\smallskip} \hline\hline \noalign{\smallskip}
Source&BP04&BP04+&BU88&Bahcall et al. 82\\
\noalign{\smallskip} \hline
pp&$-$0.084&$-$0.071&$-$0.08&$-$0.05\\
pep&$-$0.171&$-$0.147&$-$0.17&\\
hep&$-$0.242&$-$0.229&$-$0.22&\\
$^7$Be&0.619&0.637&0.58&0.60\\
$^8$B&1.364&1.369&1.27&1.26\\
$^{13}$N&1.897&1.870&1.86&1.67\\
$^{15}$O&2.056&2.038&2.03&2.00\\
$^{17}$F&2.169&2.151&2.09&\\
\noalign{\smallskip} \hline
\end{tabular}
\end{center}
\end{table}

Table~\ref{tab:alpha_iZoverX} implies that we would have made only
of order a 10\% error in the estimated total flux uncertainties
due to composition uncertainties had we used the 1982 partial
derivatives instead of their 2004 values. In all cases, using the
original 1982 partial derivatives would have changed by less than
1\% the total flux uncertainties we estimate due to composition
uncertainties.

We conclude that  the major revisions and refinements that have
been implemented in standard solar models over the past two and a
half decades have not significantly affected ${\partial \ln
\phi_i}/{\partial \ln (Z/X)}$.

\subsection{Why \boldmath${Z/X}$ changes give an overestimate of composition
changes} \label{subsec:zoverxmisleading}

The abundances of the heavier elements like Si, S, and Fe are
relatively well known and therefore these elements do not affect
greatly  $\Delta (Z/X)/ (Z/X)$, although they do affect strongly
the calculated neutrino fluxes. The influence of the heavy
elements like silicon, sulphur, magnesium, and iron on the
calculated neutrino fluxes is primarily through their affect on
the  radiative opacity in the core of the solar model.  As we
shall see in the following section (see especially
Table~\ref{tab:abundanceuncertainties}), the abundances of the
light elements like C, N, O, and Ne are relatively poorly known
and therefore they contribute a large amount to the fractional
uncertainty, $\Delta (Z/X)/ (Z/X)$. However, these light elements
do not affect the solar neutrino fluxes very much (Bahcall \&
Pinsonneault 2004); they are ionized near the base of the
convective zone, outside the region where the neutrinos are
formed.

Thus some elements, like C and O, affect strongly $\Delta (Z/X)/
(Z/X)$ without affecting very much the calculated neutrino fluxes.
Conversely, heavier elements like Fe, affect the neutrino fluxes
very significantly but, because their abundances are relatively
well known,  not $\Delta (Z/X)/ (Z/X)$.

The imperfect correlation between changes in $\Delta (Z/X)/ (Z/X)$
and changes in calculated neutrino fluxes is the basic reason that
it is necessary to evaluate individually the effects of individual
abundance uncertainties on the flux estimates.

We shall compute and present in what follows the flux uncertainties
calculated using the partial derivatives from
Table~\ref{tab:alpha_ij}, Table~\ref{tab:alpha_ij_bp04p}, and
Table~\ref{tab:alpha_iZoverX}. We will adopt the results obtained with
Table~\ref{tab:alpha_ij} for our preferred estimates of uncertainties
since the BP04 solar model is in good agreement with helioseismology
while the BP04+ solar model is in poor agreement with
helioseismological measurements in the region between a solar radius
of $R = 0.7 R_\odot$ and $R = 0.4R_\odot$ (Bahcal et al. 2005). We
also prefer the estimated uncertainties obtained with the derivatives
of Table~\ref{tab:alpha_ij} because these uncertainties are more
conservative (larger) than the estimates obtained with
Table~\ref{tab:alpha_ij_bp04p}.

\section{ABUNDANCE UNCERTAINTIES}
\label{sec:abundanceuncertainties}

In this section, we discuss and present two different estimates
for the uncertainties in the heavy element abundances on the
surface of the Sun. Our `conservative' uncertainty estimates are
obtained by comparing recent abundance analyses with previously
standard analyses. Our more `optimistic' uncertainty estimates are
obtained by adopting the uncertainty estimates in the most recent
review (Asplund et al. 2005)\nocite{asplundgrevessesauval2005}.

\subsection{Element abundances}
\label{subsec:elementabundances}

Recent analyses of the surface chemical composition of the Sun use
three-dimensional atmospheric models, take account of hydrodynamic
effects, and pay special attention to uncertainties in the atomic data
and the observed spectra.  Mass fractions that are lower than the
previous standard values (Grevesse \& Sauval 1998)\nocite{oldcomp}
have been obtained in this way for C, N, O, Ne, and Ar (Asplund et
al.~2000; \nocite{asplundetal00} Asplund 2000;\nocite{asplund00}
Allende Prieto et al. 2001, 2002;\nocite{allende01}\nocite{allende02} Asplund et
al.~2004\nocite{asplund04} ). These abundance determinations have
typical quoted uncertainties of order 0.05 dex (12\%).

However, as noted earlier the recent lower abundances of these
volatile heavy elements, when incorporated into solar models, lead
to serious discrepancies with helioseismology. And, to make the
situation even more puzzling, the solar models computed with the
previously standard heavy element abundances (Grevesse \& Sauval
1998) yield results in good agreement with helioseismology.

Estimating the uncertainty in an abundance determination is even
more difficult than arriving at a best-estimate abundance. The
important uncertainties are most often systematic and range all
the way from line blending in the observed spectra to mathematical
and physical approximations made in modeling the solar atmosphere
(see Lodders 2003\nocite{lodders03} and  Asplund et al. 2005).

We have calculated the neutrino  flux dependences on the
individual abundances of C, N, O, Ne, Mg, Si, S, Ar and Fe. These
are the elements that contribute most to the uncertainty in the
calculated fluxes.

With  the exception  of  argon, these  are  the most  abundant
metals in  the Sun. The elements not included are at least a
factor seven less abundant and their uncertainties are small
(comparable to the most precise  determination of the listed
elements).  Ar  is a special case; its abundance  is comparable to
that of other elements not listed  (for example Na, Al, Ca and Ni)
but the uncertainty in its abundance is much  larger (more than
four times) and gives  a  non-negligible contribution  to the
neutrino fluxes  uncertainties despite its low solar abundance.

The neutrino fluxes  depend on the individual metal  abundances
mainly through their  effect  in  the  radiative   opacities,  for
which  metals  are  major contributors in  the solar  interior. In
this  regard, it is  worth mentioning that for each solar model we
have produced the set of OPAL radiative opacities (using     the
tool      available       at      the       OPAL      Web
site\footnote{http://www-phys.llnl.gov/Research/OPAL/new.html})
corresponding to each of the specific compositions used.

\subsection{Conservative and optimistic abundance uncertainties}
\label{subsec:definitionabundanceuncertainties}

 We  take as one measure of the fractional uncertainty in the
heavy element abundance $\beta_i$, the difference between the
previous standard abundance minus the recently determined
abundance divided by the average of the previous abundance and the
recently determined abundance. Thus

\begin{equation}
\frac{\Delta \beta_i}{\beta_i} ~=~\frac{\left[2\left({\rm
Abundance}_{{\rm old},\,i} - {\rm Abundance}_{{\rm
new},\,i}\right)\right]}{\left[\left({\rm Abundance}_{{\rm
old},\,i} + {\rm Abundance}_{{\rm new},\, i}\right)\right]} \,\,
(1\sigma) . \label{eq:defn1sigmahistorical}
\end{equation}
Equation (\ref{eq:defn1sigmahistorical}) represents our
`conservative' estimate of abundance uncertainties and is intended
to help take account of systematic uncertainties in an empirical
way.  We use for the `old' abundances the recommended Grevesse \&
Sauval (1998) values and for the `new' abundances the values given
by Asplund et al. (2005).

We use in equation~(\ref{eq:defn1sigmahistorical}) meteoritic
abundances where available, i.e., for  Mg, Si, S, and Fe. Over
the several decades in which photospheric and meteoritic
abundances have been compared, the agreement between the
photospheric and meteoritic abundances has steadily improved.
However, when there was a conflict in the best-estimates from the
two methods, it has often turned out that the meteoritic estimate
was more accurate. We use solar abundance determinations for C, N,
O, Ne, and Ar.  The measurements for Ne and Ar are particularly
problematic and subject to systematic uncertainties (see Asplund et al. 2005).

\begin{table}[!t]
\caption{\baselineskip=16pt Adopted $1\sigma$ uncertainties for
individual heavy elements. We adopt meteoritic measurements where
available, i.e., for Mg, Si, S, and Fe (Lodders 2003).  For the
volatile elements C, N, O, Ne, and Ar we use solar atmospheric
abundances. column (2), under the heading `Historical,' gives as our
preferred (conservative) estimated error the fractional differences
between the recent abundance determinations (Asplund et al. 2005) and
the previously standard values (Grevesse \& Sauval 1998). Column (3),
'Recent Analyses' (optimistic), lists the uncertainties quoted in the
recent paper by Asplun et al. (2005); the uncertainties for meteoritic
abundances in the Asplund et al. paper are based upon Lodders (2003).
 \label{tab:abundanceuncertainties} }
\begin{center}
\begin{tabular}{ccc}
\noalign{\smallskip} \hline\hline \noalign{\smallskip}
Heavy& Historical & Recent Analyses\\
Element&[Conservative]  (\%) & [Optimistic] (\%) \\
\noalign{\smallskip} \hline
C&29.7&12.2\\
N&32.0&14.8\\
O&38.7&12.2\\
Ne&53.9&14.8\\
Mg&11.5&7.2\\
Si&11.5&4.7\\
S&9.2&9.6\\
Ar&49.6&20.2\\
Fe&11.5&7.2\\
\noalign{\smallskip} \hline
\end{tabular}
\end{center}
\end{table}

Table~\ref{tab:abundanceuncertainties} gives, in the second
column, our preferred (conservative) estimate for the fractional
uncertainties of the most important heavy elements. The
uncertainties in this column were computed using
equation~(\ref{eq:defn1sigmahistorical}) in comparing the Grevesse
\& Sauval (1998) abundances with the Asplund et al. (2005) abundances.

The third column of Table~\ref{tab:abundanceuncertainties} gives
more optimistic estimates of the composition uncertainties,
namely, the uncertainties published in the recent review by
Asplund et al. (2005).  We list the published
uncertainties without taking into account the conflict that the
new abundances cause with helioseismology and without taking
account of the previous history of the abundance determinations.
We interpret conservatively the recently published uncertainties
as $1\sigma$, although the statistical significance of the Asplund
et al. (2005) error estimates is not stated explicitly in their
paper. We use the meteoritic abundance uncertainties  whenever a
meteoritic abundance is available; interpreting the meteoritic
uncertainties as $1\sigma$ is especially conservative (Lodders
2003).

Comparing the uncertainties given in the second and third columns of
Table~\ref{tab:abundanceuncertainties}, we see that the recent
abundance determinations for the volatile elements have published
error estimates (Asplund et
al. 2005)\nocite{asplundgrevessesauval2005} that are typical one-third
of the uncertainties computed by the historical comparison expressed
by equation~(\ref{eq:defn1sigmahistorical}). The uncertainties
estimated by Asplund et al. (2005) for the meteoritic determinations
are typically a factor of one-half the uncertainties computed with
equation~(\ref{eq:defn1sigmahistorical}).

\section{UNCERTAINTIES IN INDIVIDUAL NEUTRINO FLUXES  FROM INDIVIDUAL
ELEMENT ABUNDANCES} \label{sec:individual}

We define and present in this section the uncertainties in
individual neutrino fluxes that result from the uncertainties in
each of the important element abundances. Our results are
summarized in Table~\ref{tab:individualuncertainties} and
Table~\ref{tab:optimisticuncertainties}.  In the discussion, we
point out which elements cause the largest uncertainties in the
different calculated neutrino fluxes.

The fractional uncertainty in each neutrino flux $\phi_i$ due to the
uncertainty in each element abundance can be computed to high accuracy
(Bahcall \& Ulrich 1988\nocite{bahcallulrich1988}, Bahcall
1989\nocite{book}) using
equation~(\ref{eq:definitionalphaij})--equation~(\ref{eq:powerlaws}).
Combining these equations, we can write
\begin{equation}
\frac{\Delta \phi_{i,j}}{\phi_i} ~=~  \left[1 + \frac{\Delta
\beta_j}{\beta_j} \right]^{\alpha_{ij}} ~-1 \, .
\label{eq:fractionalfluxuncertainty}
\end{equation}
The adopted fractional uncertainties ${\Delta \beta_j}/{\beta_j}$
are given in Table~\ref{tab:abundanceuncertainties} and  the
partial derivatives $\alpha_{ij}$ are given in
Table~\ref{tab:alpha_ij} and Table~\ref{tab:alpha_ij_bp04p}.

Table~\ref{tab:individualuncertainties} presents our best
estimates of the uncertainties in each solar neutrino flux due to
the uncertainty in each element abundance. The results given in
Table~\ref{tab:individualuncertainties} were evaluated using the
partial derivatives computed for the BP04 solar model and given in
Table~\ref{tab:alpha_ij};   the conservative composition
uncertainties are given in the second column of
Table~\ref{tab:abundanceuncertainties}.

The abundance of iron contributes the largest uncertainty from
element abundances for both the $^7$Be and the $^8$B solar
neutrino fluxes. Neon  is a close second, with oxygen making the
third largest contribution.

\begin{table}[!t]
\small \caption{\baselineskip=16pt Uncertainties of individual
neutrino fluxes from individual heavy element abundances. The
entries in the table are the fractional uncertainties, $\Delta
\phi_{i,j}/\phi_i$, of each of the principal solar neutrino
fluxes, due to each of the heavy element abundances, $\beta_j$,
defined by equation~(\ref{eq:definitionbeta}). The partial
derivatives that were used were computed with the aid of the BP04
solar model with the standard composition from Grevesse \& Sauval
(1998) and are listed in Table~\ref{tab:alpha_ij}. The adopted
uncertainties for each heavy element are given in column (2) of
Table~\ref{tab:abundanceuncertainties} under the label `Historical
[Conservative].' \label{tab:individualuncertainties} }
\begin{tabular}{@{\extracolsep{-7pt}}lrrrrrrrrr}
\noalign{\smallskip} \hline\hline \noalign{\smallskip}
Source&C   &N&O&Ne&Mg&Si&S&Ar&Fe\\
\noalign{\smallskip} \hline
pp&$-$3.63E-3&$-$8.33E-4&$-$1.96E-3&$-$2.15E-3&$-$5.44E-4&$-$1.20E-3&$-$7.04E-4&$-$8.05E-4&$-$2.50E-3\\
pep&$-$6.48E-3&$-$1.66E-3&$-$3.59E-3&$-$2.15E-3&$-$5.44E-4&$-$1.52E-3&$-$1.50E-3&$-$2.41E-3&$-$7.05E-3\\
hep&$-$3.89E-3&$-$1.11E-3&$-$7.50E-3&$-$7.30E-3&$-$1.96E-3&$-$4.02E-3&$-$2.46E-3&$-$2.82E-3&$-$7.48E-3\\
$^7$Be&$-$5.20E-4&5.55E-4&1.72E-2&2.14E-2&5.57E-3&1.14E-2&6.53E-3&7.28E-3&2.30E-2\\
$^8$B&7.83E-3&3.06E-3&4.04E-2&4.23E-2&1.05E-2&2.13E-2&1.21E-2&1.38E-2&5.77E-2\\
$^{13}$N&2.46E-1&5.15E-2&2.62E-2&2.49E-2&6.55E-3&1.40E-2&8.31E-3&9.71E-3&3.79E-2\\
$^{15}$O&2.40E-1&5.97E-2&3.09E-2&2.98E-2&7.65E-3&1.65E-2&9.64E-3&1.13E-2&4.46E-2\\
$^{17}$F&8.62E-3&2.78E-3&4.34E-1&3.33E-2&8.53E-3&1.80E-2&1.06E-2&1.26E-02&4.95E-2\\
\noalign{\smallskip} \hline
\end{tabular}
\end{table}

The iron abundance is relatively well determined (see
Table~\ref{tab:abundanceuncertainties}) from both photospheric and
meteoritic measurements (see Asplund et al. 2005 and Lodders
2003). Nevertheless, the uncertainty from iron is large because of
the relatively large partial derivatives of the fluxes with
respect to the iron over hydrogen abundance ratio (see
Table~\ref{tab:alpha_ij}).  The oxygen abundance has recently been
the subject of very comprehensive studies using the solar
atmospheric spectrum. However, it is notoriously difficult to
measure the abundance of the noble gases neon and argon, since
they are absent in the Fraunhofer absorption line spectrum of the
solar photosphere and are mostly lost from meteorites.
Measurements of the neon and argon abundance must be made
indirectly using coronal data, the solar wind, and energetic solar
particles. Since we have limited knowledge of the details of the
physics within the regions where neon lines are formed, these
indirect measurements are subject to unknown systematic errors.

The abundance of carbon contributes the largest composition
uncertainty for the p-p, $^{13}$N, and $^{15}$O solar neutrino
fluxes. In fact, the uncertainty in the carbon abundance dominates
the composition uncertainty for the $^{13}$N and $^{15}$O neutrino
fluxes.

Table~\ref{tab:optimisticuncertainties} presents more optimistic
estimates for the uncertainties in the individual neutrino fluxes
from individual heavy elements. These smaller uncertainties are
calculated using the abundance uncertainties given by Asplund et
al. (2005) \hbox{(column [3] of
Table~\ref{tab:abundanceuncertainties})} and the partial
derivatives computed using the BP04+ solar model
(Table~\ref{tab:alpha_ij_bp04p}).

\begin{table}[!t]
\small \caption{\baselineskip=16pt {Optimistic uncertainties of
individual neutrino fluxes from individual heavy element
abundances, ${\phi_{i,j}/\phi_i}$.} The present table is similar
to  Table~\ref{tab:individualuncertainties}. However, for the
present table we used the more optimistic uncertainties estimated
by Asplund et al. (2005); these optimistic uncertainties are
listed in column (3) of Table~\ref{tab:abundanceuncertainties}
under the label `Recent Analyses [Optimistic].'  Also, we have
used here the partial derivatives computed for the solar model
BP04+. \label{tab:optimisticuncertainties} }
\begin{tabular}{@{\extracolsep{-7pt}}lrrrrrrrrr}
\noalign{\smallskip} \hline\hline \noalign{\smallskip}
Source&C&N&O&Ne&Mg&Si&S&Ar&Fe\\
\noalign{\smallskip} \hline
pp&$-$1.15E-3&$-$4.14E-4&$-$5.75E-4&$-$4.14E-4&$-$3.48E-4&$-$4.59E-4&$-$6.41E-4&$-$1.84E-4&$-$1.53E-3\\
pep&$-$2.07E-3&$-$5.52E-4&$-$9.20E-4&$-$2.76E-4&$-$2.09E-4&$-$6.89E-4&$-$1.37E-3&$-$5.52E-4&$-$4.30E-3\\
hep&$-$1.38E-3&$-$4.14E-4&$-$2.07E-3&$-$1.52E-3&$-$1.32E-3&$-$1.79E-3&$-$2.65E-3&$-$9.20E-4&$-$4.99E-3\\
$^7$Be&5.76E-4&2.76E-4&5.31E-3&4.57E-3&3.97E-3&5.30E-3&7.36E-3&2.21E-3&1.61E-2\\
$^8$B&4.04E-3&1.24E-3&1.15E-2&8.87E-3&7.47E-3&9.78E-3&1.38E-2&4.24E-3&3.92E-2\\
$^{13}$N&1.02E-1&2.52E-2&6.35E-3&4.98E-3&4.53E-3&6.40E-3&9.39E-3&2.76E-3&2.50E-2\\
$^{15}$O&9.95E-2&2.96E-2&7.86E-3&5.95E-3&5.37E-3&7.51E-3&1.10E-2&3.32E-3&2.98E-2\\
$^{17}$F&4.04E-3&1.10E-3&1.32E-1&6.65E-3&6.00E-3&8.30E-3&1.20E-2&3.69E-3&3.31E-2\\
\noalign{\smallskip} \hline
\end{tabular}
\end{table}

The largest composition uncertainty in the calculation of the
$^7$Be and $^8$B neutrino fluxes is again the iron abundance given
the assumptions used in calculating
Table~\ref{tab:optimisticuncertainties}, just as it was for the
more conservative uncertainty estimates used in constructing
Table~\ref{tab:individualuncertainties}. However, for the
assumptions used in calculating
Table~\ref{tab:optimisticuncertainties}, sulphur contributes the
second largest uncertainty and oxygen, neon, and silicon all
contribute significantly.

 For the p-p neutrino flux, the largest  uncertainty due to
composition is from the iron abundance (for the conditions of
Table~\ref{tab:optimisticuncertainties}). But, for the $^{13}$N
and $^{15}$O neutrino fluxes, the carbon abundance dominates the
uncertainty due to composition.

Why do we use for Table~\ref{tab:optimisticuncertainties} the partial
derivatives computed with the BP04+ model?  Since we use in
Table~\ref{tab:optimisticuncertainties} the uncertainties determined
by recent measurements, it is more appropriate to use the partial
derivatives that are obtained with a model, BP04+, that adopts the
recent abundance determinations. The precise composition used in
constructing BP04+ is given in Table~1 of Bahcall et
al. (2004b). There are slight differences between the composition used
in constructing BP04+ and the very latest estimated abundances given
by Asplund et al. (2005). To make sure that these small differences
were unimportant, we evolved a solar model that was identical to BP04
and BP04+ except that the new model, BP04AGS, uses the Asplund et
al. (2005) abundances. The average difference between the total
neutrino flux uncertainties due to composition uncertainties that was
obtained using models BP04+ and BP04AGS, was 0.09\% and, in all cases,
less than 0.2\%.

\section{NEUTRINO FLUX AND EXPERIMENTAL RATE UNCERTAINTIES FROM
ALL ABUNDANCE UNCERTAINTIES}
\label{sec:allcompositionuncertainties}

We begin this section by describing in
\S~\ref{subsec:totalabundanceuncertainties} how we combine all of
the abundance uncertainties to compute the total uncertainty in
the calculated neutrino fluxes that arise from all composition
determinations, assuming that the abundance determinations of
different elements are statistically independent. We discuss in
\S~\ref{subsec:correlation} the special case of the neon and argon
abundances, which are determined relative to a reference element
that can be measured in the solar photosphere. We discuss in
\S\ref{subsec:correlationsmeteoritic} the correlated uncertainty
in the meteoritic abundances that results from adjusting the
meteoritic abundance scale to agree with the solar atmospheric
abundance scale. We compare in
\S~\ref{subsec:conservativevsoptimistic} the conservative
uncertainty estimates (our preferred estimates) with the more
optimistic uncertainty estimates that are based upon the published
error estimates of abundance uncertainties by Asplund et al.
(2005). In \S~\ref{subsec:zoverxuncertainties}, we compare the
uncertainties that have been traditionally estimated using the
total $Z/X$ with the uncertainties that are estimated using
individual abundance uncertainties.

Table~\ref{tab:totalcompositionuncertainties}  summarizes the
principal results of this section.

The software used to combine the abundance uncertainties is
available at \\ http://www.sns.ias.edu/$\sim$jnb under the menu items
Solar Neutrinos/\hbox{software and data}. The code,
exportrates.f, provides options for calculating the uncertainties
using individual abundance uncertainties and also using the
uncertainty in the total $Z/X$.

\subsection{Computation of flux uncertainties from all abundance
uncertainties}
\label{subsec:totalabundanceuncertainties}

 The uncertainty due to abundance determinations for a given calculated neutrino
flux can be obtained by combining the effects of all the element
uncertainties on the flux of interest. If all of the abundance
determinations are independent of each other, the uncertainties
can be combined quadratically. However, if there are correlations
in the uncertainties in the abundance determinations, these must
be taken into account. Usually, observers do not specify the
correlations among the quoted  uncertainties. However, as we
discuss in \S~\ref{subsec:correlation},  there is a strong
correlation between the uncertainties in the neon and argon
abundances and in the oxygen abundance and we  take this into
account.

The general formula for the uncertainty in the neutrino flux
$\phi_i$ can be written
\begin{equation}
\frac{\Delta \phi_{i}}{\phi_i}~=~\sqrt{\sum_j \left( \frac{\Delta
\phi_{i,j}}{\phi_i}  \right)^2 + \sum_{k \neq l} \left(
\frac{\Delta \phi_{i,k}}{\phi_i}\right) \left( \frac{\Delta
\phi_{i,l}}{\phi_i}\right)\rho(k,l)}\, ,
\label{eq:totalfractionalfluxuncertainty}
\end{equation}
where the indices \hbox{$j, k$, and $l$} denote different elements and
$\rho(k,l)$ is the correlation coefficient between the abundance
uncertainties of the \hbox{$k$ and $l$} elements. If the uncertainties
of two elements, $k$ and $l$, are uncorrelated then $\rho(k,l) =
0.0$.  If the uncertainties are fully correlated, $\rho(k,l) =
1.0$.

 For radiochemical experiments, chlorine, gallium, and lithium
experiments, the procedure for calculating the uncertainty,
$\Delta R$, in predicted event rates, $R$,  is somewhat more
complicated. The measured rates for radiochemical experiments are
sensitive to contributions from different neutrino branches, with
each neutrino flux contributing an amount $\phi_i \sigma_i$ (where
$\sigma_i$ is the neutrino absorption cross section). The
uncertainty from a given element abundance, $\beta_j$, affects in
a coherent way to all of the partial contributions $\phi_i
\sigma_i$. The contribution to the uncertainty in the rate from a
fixed composition $j$, $\Delta r(j)$, is
\begin{equation}
\Delta r(j) = \sum_i \phi(i) \sigma(i) \left( \frac{\Delta
\phi_{i,j}}{\phi_i}\right) \, .
\label{eq:partialrateuncertainty}
\end{equation}
To calculate the uncertainty for a radiochemical rate with only
incoherent contributions, we first sum over all partial rate
contributions $i$ for a fixed composition uncertainty $j$, before
quadratically combining the uncertainties from each different
composition $j$.

If the uncertainties from some elements are correlated, then we
must include the effects of the correlations as an additional
term. Including the possibility of correlations, the general
expression for the uncertainty in a radiochemical experiment is
\begin{equation}
\Delta {\rm Rate} ~=~  \sqrt{\sum_j \left( \Delta r(j)\right)^2 +
\sum_{k \neq l}  \Delta r(k)  \Delta r(l)\rho(k,l)}\, .
\label{eq:radiochemicalfractionaluncertainty}
\end{equation}

\subsection{Correlation of neon, argon, and oxygen abundance
uncertainties}
\label{subsec:correlation}

The noble gases neon and argon do not appear in the solar
photospheric spectrum and are largely lost by meteorites.
Therefore, the abundances of neon and argon abundances must be
determined in environments that are less well understood than the
photospheric spectrum, in particular, in the coronal spectrum, in
the solar wind, in solar energetic particles, and by gamma ray
spectroscopy. The measurements  of the neon and argon abundances
must be  made with respect to some reference element that does
appear in the solar photosphere.  The reference element of choice
is usually oxygen or magnesium. There are, of course, potentially
very large systematic uncertainties in these indirect
determinations of the neon and argon abundances.

Asplund et al. (2005) give neon and argon abundances that are
determined relative to the oxygen abundance. Thus the
uncertainties in the neon and argon abundances are correlated with
the uncertainty in the oxygen abundance. From the uncertainties
given in Table~1 of Asplund et al. (2005), we infer that Asplund
et al. believe that the uncertainty in the neon abundance is
dominated by the uncertainty in the oxygen abundance and that the
uncertainty in the argon abundance is due to comparable
contributions from the measurement of the oxygen abundance and
from the Ar/O ratio.

We have made calculations based upon two extreme assumptions.
First, we assume that all the abundance determinations, including
those of oxygen, neon, and argon, are independent. Second, we
assume that the uncertainties in the oxygen, neon, and argon
abundances are completely correlated. Given this second
assumption,   we set $\rho(k,l)= 1.0$ in
equation~(\ref{eq:totalfractionalfluxuncertainty}) and
equation~(\ref{eq:radiochemicalfractionaluncertainty}) when both
$k$ and $l$ represent either O, Ne, or Ar (otherwise, we take
$\rho(k,l)= 0.0$).

 Fortunately, the differences in the flux uncertainties are
relatively small when the uncertainties are calculated in these
two extreme ways. The average fractional difference in a neutrino
flux uncertainty due to assuming complete correlation (of the O,
Ne, and Ar uncertainties) or no correlation varies between 6\% and
17\% of the composition uncertainty itself, depending upon which
partial derivatives (BP04 or BP04+) are used and depending upon
whether we adopt conservative or optimistic uncertainties for the
element abundances. Since the composition uncertainty is only one
of a number of different sources of flux uncertainty (see
discussion in \S~\ref{sec:alluncertainties}), a 17\% uncertainty
in the composition uncertainty  is acceptable. We certainly do not
believe the composition uncertainties discussed in
\S~\ref{sec:abundanceuncertainties} and
Table~\ref{tab:abundanceuncertainties} are reliable to 17\% of the
quoted uncertainty. After all, the observers do not as a rule
specify the confidence level which their uncertainties represent
(see, however, the Lodders 2003 analysis of meteoritic
abundances).

In compiling Table~\ref{tab:totalcompositionuncertainties}, we
again made a conservative assumption, namely, that the oxygen,
neon, and argon uncertainties are completely correlated.  Since
all three elements have, for a given neutrino flux,  the same sign
for their partial derivatives (see Table~\ref{tab:alpha_ij} and
Table~\ref{tab:alpha_ij_bp04p}), this procedure results in a
larger uncertainty than if we had combined incoherently the
oxygen, neon, and argon uncertainties. The assumption we make here
slightly overestimates the total composition uncertainty assuming
the correctness of all the other numbers that go into the
analysis.

In future compilations of solar abundances and their
uncertainties, it will be very useful if the compilers specify the
correlation between the difference abundance uncertainties.

\subsection{Correlations of meteoritic abundances via a scale
factor}
\label{subsec:correlationsmeteoritic}

Traditionally, the solar abundances of the elements that can be
measured in the solar photosphere or elsewhere in the solar
atmosphere are determined relative to the abundance of hydrogen.
However, hydrogen is lost from meteorites. Therefore, the
abundances that are measured in meteorites are determined relative
to some other element, which is usually taken as Si. The two
scales, the atmospheric and the meteoritic, are adjusted to give a
consistent set of values by sliding one of the scales up or down
with respect to the other scale. For specificity, we can think of
this procedure as adjusting the meteoritic scale with respect to
the atmospheric scale.

The uniform adjustment of the meteoritic scale implies that there
is a correlation of all of the meteoritic abundances among
themselves. The amount of this correlation can be estimated by
calculating how much we have to change the meteoritic scale with
respect to the atmospheric scale in order to significantly affect
the goodness of the agreement between the two scales, which is
generally excellent (see Grevesse \& Sauval 1998, Lodders 2003,
Asplund et al. 2005).  We therefore need to
evaluate
\begin{equation}
\sigma_{\rm{meteorite,~atmosphere}}~=~\sqrt{\frac{1}{N}\sum_i
\left(\beta_{{\rm meteorite,}i} - \beta_{{\rm
atmosphere},i}\right)^2} \, ,
\label{eq:meteoriteatmosphere}
\end{equation}
where $\beta_{{\rm meteorite,}i}$ and  $\beta_{{\rm
atmosphere},i}$ are, respectively, the meteoritic and atmospheric
abundances of the element $i$.

We are interested in the accuracy with which the meteoritic and
atmospheric scales can be brought into agreement for the abundant
heavy elements, since it is only the abundant heavy elements that
affect the neutrino fluxes significantly through their
contributions to the radiative opacity. Moreover, it is plausible
that the relative abundances are more robust for elements that are
more abundant.

We have evaluated the correlation that is expressed in
equation~(\ref{eq:meteoriteatmosphere}) for a number of different
cases. We find that the correlation is small and the general size
of the correlation is robust. For example, we have computed the
correlation $\sigma_{\rm{meteorite,~atmosphere}}$ for the seven
elements with logarithmic abundances relative to hydrogen that are
greater than 6.0 on the usual scale in which the hydrogen
abundance is set equal to 12.0 .  For the Lodders (2003)
abundances, the unweighted average is
$\sigma_{\rm{meteorite,~atmosphere}} = 0.7\%$ and the weighted
average (quadratically combined atmospheric and meteoritic errors)
is $\sigma_{\rm{meteorite,~atmosphere}} = 0.9\%$. These results
are essentially unchanged if we throw out, e.g., the Ni abundance,
which is the least well determined of the sample we are
considering. Similar results are obtained for the Asplund et al.
(2005) abundances. We find an unweighted average of
$\sigma_{\rm{meteorite,~atmosphere}} = 1.8\%$ and a weighted
average of 2.1\%.

The correlated contribution of the uncertainty due to the relative
adjustment of the meteoritic and atmospheric scales depends
quadratically upon $\sigma_{\rm{meteorite,~atmosphere}}$. It
follows from the definition of the correlation coefficient
$\rho(k,l)$ between the uncertainties of the meteoritic abundances
of two elements $k \neq l$ that
\begin{equation}
\rho(k,l) ~=~\frac{\sigma_{k,l}}{\sigma_k \sigma_l} \,
\label{eq:rhoisquadratic}
\end{equation}
where $\sigma_{k,l}$ is the covariance between the uncertainties
for the two elements and $\sigma_k$ and $\sigma_l$ are the
uncertainties of each element abundance. The individual
uncertainties, $\sigma_k$ and $\sigma_l$, include the
quadratically-combined meteoritic measurement uncertainty plus the
covariance from the scale adjustment.

We do not have a good way of calculating the covariance, but we
hope that the covariance will be evaluated in the future by the
authors of papers presenting critical summaries of solar element
abundances. We can obtain a reasonable upper limit to the
covariance by assuming that it is, in order of magnitude, the
square of $\sigma_{\rm{meteorite,~atmosphere}}$ (see
eq.~[\ref{eq:meteoriteatmosphere}]).

The precise values of $\rho(k,l)$ depend upon whose compilation of
solar abundances one uses and upon how one defines the sample of
abundant heavy elements within the published list of abundances.
However, in order of magnitude,
 $\sigma_{\rm atmospheric,} \sim \sigma_{\rm meteoritic} \sim
0.1$ and the characteristic value of $\sigma_{k,l} \sim 0.01$.
Hence, $\rho(k,l) \sim 0.01$.

Since $\rho$ is small, we can neglect, without making a
significant numerical error, the effects of the relative
adjustment of the meteoritic and atmospheric abundance scales in
equation~(\ref{eq:totalfractionalfluxuncertainty}) and
equation~(\ref{eq:radiochemicalfractionaluncertainty}). This
approximation is satisfactory even though there are seven times as
many off-diagonal terms as diagonal terms, .

In the future, when there is general agreement on the correct solar
abundances and their uncertainties, the effect of the relative
adjustment of the meteoritic and atmospheric scales can be taken into
account using equations~(\ref{eq:totalfractionalfluxuncertainty}),
(\ref{eq:radiochemicalfractionaluncertainty}), and
(\ref{eq:meteoriteatmosphere}).

\subsection{Comparison of conservative and optimistic uncertainty
estimates}
\label{subsec:conservativevsoptimistic}

Table~\ref{tab:totalcompositionuncertainties} presents our
estimates for the total uncertainty in each calculated neutrino
flux and in each radiochemical rate due to all composition
sources. We present in the table the results from a conservative
estimate (column [2]), an optimistic estimate (column [3]), and the
traditional method that uses the total $Z/X$. The entries in the
table were computed using
equation~(\ref{eq:totalfractionalfluxuncertainty}) and
equation~(\ref{eq:radiochemicalfractionaluncertainty}), except
that we have combined coherently the uncertainties in the oxygen,
neon, and argon abundances as described in
\S~\ref{subsec:correlation}.

The second column of Table~\ref{tab:totalcompositionuncertainties}
presents our best estimate for the total uncertainty from all
composition uncertainties for each neutrino flux and for the rate
of each radiochemical experiment.  The partial derivatives used in
these calculations were taken from Table~\ref{tab:alpha_ij} (solar
model BP04)  and, for the entries enclosed in parentheses,
Table~\ref{tab:alpha_ij_bp04p} (solar model BP04+). We also used
for the second column of
Table~\ref{tab:totalcompositionuncertainties} the conservative
individual abundance uncertainties listed in the second column of
Table~\ref{tab:abundanceuncertainties}.

We present the much more optimistic estimates of the total
composition uncertainties in the third column of
Table~\ref{tab:totalcompositionuncertainties}. The entries in the
third column were calculated using the  abundance uncertainties
(Asplund et al. 2005) that are listed in the third
column of Table~\ref{tab:abundanceuncertainties}. We also used the
BP04+ partial derivatives from Table~\ref{tab:alpha_ij_bp04p}.

\begin{table}[!t]
\caption{\baselineskip=12pt Neutrino Flux and Rate Uncertainties
from All Solar Composition Uncertainties.   The abundance
uncertainties used in calculating the entries in the second column
of the present table were obtained  using the historical
(conservative) composition uncertainties that are listed in the
second column of Table~\ref{tab:abundanceuncertainties}; the
uncertainties from recent analyses (our `optimistic'
uncertainties) are given in the third column of
Table~\ref{tab:abundanceuncertainties} and were used to obtain the
entries in the third column of the present table. The
uncertainties without parentheses that are listed in columns (2) and
(3) of the table were calculated using partial derivatives from
Table~\ref{tab:alpha_ij} that were obtained with solar models that
had the BP04, i.e., the 1998 Grevesse \& Sauval solar composition
of heavy elements. The uncertainties in parentheses that are
listed in columns (2) and (3) were calculated with partial derivatives
from Table~\ref{tab:alpha_ij_bp04p} that were obtained with solar
models that had the BP04+ recently-determined solar composition
(see Table~1 of Bahcall, Serenelli, \& Pinsonneault 2004). The
uncertainties in column (4) were calculated assuming that (see
Bahcall \& Pinsonneault 2004) the total spread in all modern
measurements of the heavy element abundance by mass divided by the
hydrogen abundance by mass, $Z/X$, is equal to the $3\sigma$
uncertainty in $Z/X$,\label{tab:totalcompositionuncertainties} i.e., ${\Delta (Z/X)}/{(Z/X)} = 0.15 (1\sigma)$. For column (4), the
values without parentheses were calculated by Bahcall \&
Pinsonneault (2004) with the Bahcall \& Ulrich (1988) partial
derivatives and the values with parentheses were calculated with
the BP04 partial derivatives (see Table~\ref{tab:alpha_iZoverX}).}
\begin{center}
\begin{tabular}{lccc}
\noalign{\smallskip} \hline\hline \noalign{\smallskip}
Neutrino&Historical& Recent Analyses&$Z/X$:\\[-1pt]
Flux& (Conservative)  (\%) & (Optimistic)(\%) & Historical (\%)\\[-1pt]
\noalign{\smallskip} \hline\noalign{\smallskip}
pp&0.7 (0.5)&0.3(0.2)&1.0 (1.0)\\[-1pt]
pep&1.3 (1.0) &0.6(0.5) &2.0 (2.1)\\[-1pt]
hep&2.0(1.6)&0.9(0.8)&2.6 (2.9)\\[-1pt]
$^7$Be&5.3 (4.6)&2.4(2.2)&8.0 (8.6)\\[-1pt]
$^8$B&11.6 (9.9) &5.3(5.0)&20.0 (22.0)\\[-1pt]
$^{13}$N&26.2 (25.8)&11.1(11.0)&33.2 (34.1)\\[-1pt]
$^{15}$O&26.2 (25.7)&11.2(11.0)&37.5 (38.2)\\[-1pt]
$^{17}$F&48.3 (45.4)&15.6(14.7)&39.1 (41.3)\\[-1pt]
\noalign{\smallskip} \hline\noalign{\smallskip}
Experiment&SNU&SNU&SNU\\
\noalign{\smallskip} \hline \noalign{\smallskip}
$^{37}$Cl& 0.9(0.8)&0.4(0.4)&1.6 (1.7)\\[-1pt]
$^{71}$Ga& 4.4(4.0)&1.9(1.9)&8.2 (8.7)\\[-1pt]
$^{7}$Li&5.7(5.3)&2.5(2.5)&10.6 (11.2) \\[-1pt]
\noalign{\smallskip} \hline
\end{tabular}
\end{center}
\end{table}

The conservative composition uncertainties listed in the second
column of Table~\ref{tab:totalcompositionuncertainties} are
typically a factor of two or more larger than the more optimistic
composition uncertainties listed in the third column of the table.

\subsection{Uncertainties calculated using total \boldmath$Z/X$}
\label{subsec:zoverxuncertainties}

The last (fourth) column of
Table~\ref{tab:totalcompositionuncertainties} lists the
composition uncertainties that are computed by using the
traditional method of lumping together all heavy elements, i.e.,
using the partial derivatives $\partial \ln \phi_i/\partial \ln
(Z/X)$ (see eq.~[\ref{eq:definitionalphaiZoverX}]). In this
calculation, we assumed that $\Delta (Z/X)/(Z/X) = 0.15
(1\sigma)$, based upon the historical time dependence of the
published values of $Z/X$ (Bahcall \& Pinsonneault 2004).

The  traditional total $Z/X$, or historical, method (last column
of Table~\ref{tab:totalcompositionuncertainties}) yields (except
for the C, N, O neutrino fluxes) composition uncertainties that
vary from about 50\% larger to  a factor of two larger than the
conservative uncertainties (column (2) of
Table~\ref{tab:totalcompositionuncertainties}) estimated by our
preferred method that is based upon individual abundances (second
column of Table~\ref{tab:totalcompositionuncertainties} ).  The
two most important special cases are the $^7$Be and the $^8$B
neutrino fluxes for which the traditional $Z/X$ method gave
uncertainties of 8.0\% and 20.0\%, respectively.  Our best
estimates (conservative)using errors on individual abundance
determination are 5.3\% and 11.6\% for these same two neutrino
fluxes. For the CNO neutrino fluxes, $^{13}$N, $^{15}$O, and
$^{17}$F, the estimated total uncertainties are comparable for
both ways of computing the total uncertainties (individual
abundances and a single $Z/X$).

The total composition uncertainties for the $^{37}$Cl, $^{71}$Ga,
and $^7$Li radiochemical solar neutrino experiments are reduced by
about a factor of two by taking account of the sensitivities to
individual compositions.

\section{UNCERTAINTIES FROM ALL KNOWN SOURCES}
\label{sec:alluncertainties}

We present in \S~\ref{subsec:totalresults} our estimates for the
total uncertainties in the solar neutrino fluxes calculated with
standard solar models. We present uncertainties obtained with our
preferred conservative approach, with a more optimistic approach,
and with the traditional $Z/X$ approach. We compare and discuss in
\S~\ref{subsec:comparisontotal} the total uncertainties obtained
by different methods.

\begin{table}[!t]
\caption{Neutrino Flux Uncertainties from all
known sources. The uncertainties listed in the table  include all
known sources of uncertainties (see Bahcall \& Pinsonneault
2004). The labels of the columns have the same meaning as in
Table~\ref{tab:totalcompositionuncertainties}. Columns (2)--(4) differ
only in the way the uncertainties due to the solar heavy element
abundances were calculated. However, each entry in the present
table includes the quadratically combined total uncertainties from
all sources rather than just the uncertainties from the
composition as given in
Table~\ref{tab:totalcompositionuncertainties}.
\label{tab:totaluncertainties} }
\begin{center}
\begin{tabular}{lccc}
\noalign{\smallskip} \hline\hline \noalign{\smallskip}
Neutrino&Historical&Recent Analyses&$Z/X$:\\
Flux&(conservative)   (\%) & (optimistic) (\%) & Historical (\%)\\
\noalign{\smallskip} \hline
pp&1.0 (0.9)&0.8(0.8)&1.2 (1.3)\\
pep&1.7 (1.5) &1.3 (1.2) &2.3(2.3)\\
hep&15.5 (15.5)&15.4 (15.4)&15.6 (15.6)\\
$^7$Be&10.5 (10.1)&9.3 (9.3)&12.1 (12.5)\\
$^8$B&16.3 (15.1) &12.6 (12.5)&23.0 (24.8)\\[3pt]
$^{13}$N&$^{+31.2}_{-28.1}$($^{+30.9}_{-27.8}$)&$^{+20.2}_{-15.1}$($^{+20.2}_{-15.1}$)&$^{+37.3}_{-34.8}$ ($^{+38.1}_{-35.6}$)\\[3pt]
$^{15}$O&$^{+33.2}_{-28.8}$ ($^{+32.8}_{-28.4}$)&$^{+23.3}_{-16.4}$($^{+23.2}_{-16.4}$)&$^{+42.7}_{-39.4}$($^{+43.3}_{-40.1}$)\\[3pt]
$^{17}$F&52.2 (49.5)&25.1 (24.5)&43.8 (45.7)\\
\noalign{\smallskip} \hline
Experiment&SNU&SNU&SNU\\
\noalign{\smallskip} \hline
$^{37}$Cl& 1.3( 1.2)& 1.0 ( 1.0)& 1.8 (1.9)\\[3pt]
$^{71}$Ga& {$^{+9.5}_{-9.5}$}($^{+9.4}_{-9.3}$)& {$^{+8.7}_{-8.6}$}($^{+8.7}_{-8.6}$)&$^{+11.8}_{-11.7}$ ($^{+12.2}_{-12.1}$) \\[3pt]
$^{7}$Li&$^{+7.6}_{-7.2}$($^{+7.3}_{-7.0}$)& $^{+5.6}_{-5.1}$($^{+5.6}_{-5.1}$)& $^{+11.7}_{-11.5}$($^{+12.3}_{-12.1}$)\\[3pt]
\noalign{\smallskip} \hline
\end{tabular}
\end{center}
\end{table}

\subsection{Total uncertainties in neutrino fluxes from all sources}
\label{subsec:totalresults}

The uncertainties in neutrino fluxes and rates due to  all the
known sources of uncertainties, including nuclear reaction rates,
radiative opacity, element diffusion, the solar luminosity as well
as element abundances, can be computed using
equation~(\ref{eq:totalfractionalfluxuncertainty}) and
equation~(\ref{eq:radiochemicalfractionaluncertainty}). Each separate
source of uncertainty is represented by an index $j$ in these
equations. We adopt the uncertainties for all of the sources
except element abundances as presented in the recent discussion by
Bahcall \& Pinsonneault (2004).

The uncertainties due to  radiative opacity and element diffusion
deserve special attention, since they could in principle be
affected by the recent abundance determiantions. The uncertainty
due to radiative opacity has been calculated by comparing the
neutrino fluxes computed with solar models that used the older Los
Alamos and the the much-improved and (then) new Livermore
opacities, the first and third rows of Table~7 of Bahcall and
Pinsonneault (1992)\nocite{BP92}. The fractional differences in
the calculated neutrino fluxes were taken to be $3\sigma$.  We
expect that the estimated uncertainties due to radiative opacities
will be reduced somewhat when a similar comparison is made between
the Livermore opacities and the recently published OPAL opacities
(Badnell et al. 2004, Seaton \& Badnell et al. 2004).

The uncertainties in the neutrino fluxes due to element diffusion
are calculated in three steps. First, a solar model was
constructed with full element diffusion and then a similar model
was constructed with no diffusion (see Model~9 and Model~10
Table~III of Bahcall \& Pinsonneault 1995)\nocite{BP95}.  Next,
the fractional differences in the neutrino fluxes between the two
models are formed. These differences represent the extreme changes
between no diffusion and our best estimate for diffusion. Finally,
the fractional differences are multiplied by 15\%, which is the
$3\sigma$  error estimate on the calculation of the diffusion rate
made in the original paper of Thoul, Bahcall, and Loeb
(1994)\nocite{Thoul94}\footnote{We calculated solar models with
the Asplund et al. (2005) composition but with a diffusion
coefficient changed by $\pm 15$\% from the best-estimate of Thoul
et al. (1994). The many-sigma disagreements with the
helioseismology persist.  There are moderate improvements in the
agreement with some of the quantities but moderate increases in
the disagreements with other quantities. We conclude that
adjustments in the diffusion coefficient can not resolve the
discrepancy between solar model predictions and helioseismological
measurements if the Asplund et al. (2005) abundances are adopted.}

Table~\ref{tab:totaluncertainties} presents the total
uncertainties from all sources. The structure of
Table~\ref{tab:totaluncertainties} is similar to
Table~\ref{tab:totalcompositionuncertainties}.  For both tables,
the column labeled `Historical' was calculated by interpreting as
$1\sigma$ uncertainties the difference between the best-estimates
for element abundances given by Grevesse \& Sauval (1998) and by
Asplund et al. (2005). For the third column, labeled `Recent
Analyses,' we adopted the abundance uncertainties given in the
recent review by Asplund et al. (2005). The last column, labeled
`$Z/X$: Historical' assumes that $\Delta (Z/X)/(Z/X) = 0.15 (1
\sigma)$. The entries in columns (2) and (3) without parentheses
(in parentheses) were calculated using logarithmic partial
derivatives with respect to the BP04 (BP04+) solar models. The
entries without parentheses in the last column, `$Z/X$:
Historical,' were calculated by Bahcall \& Pinsonneault (2004)
using the logarithmic partial derivatives of Bahcall \& Ulrich
(1988); the entries in parentheses in column (4) were calculated
with partial derivatives obtained using the BP04 solar model.

\subsection{Comparison of total uncertainties obtained with different
assumptions}
\label{subsec:comparisontotal}

Our preferred, most conservative estimates are given in column (2)
of Table~\ref{tab:totaluncertainties} and should be compared with
the more optimistic estimates of uncertainties given in column
(3). The total uncertainty in the $^8$B neutrino flux is most
affected by the difference between the two methods of calculating
the composition uncertainties.  The conservative approach yields a
$16.3$\% uncertainty for the calculated $^8$B solar neutrino flux,
while the optimistic approach yields  a 12.5\% uncertainty, which
is 30\% smaller.

The traditional (total $Z/X$) method of estimating the uncertainty
from the solar composition yields a much larger value for the
total uncertainty, 23.0\% (24.8\% with BP04 partial derivatives
used for both cases). The $Z/X$ method therefore overestimates the
total uncertainty from all sources by 41\% (51\% using BP04
partial derivatives for both cases) and by a factor of two
relative to the optimistic individual composition uncertainties.

The $^8$B neutrino flux is extremely rare but also extremely
important. The SNO (Ahmed et al. 2004)\nocite{snosalt} and
Super-Kamiokande (Fukuda et al. 2001)\nocite{fukuda01}  solar
neutrino experiments measure only $^8$B neutrinos and the chlorine
radiochemical experiment (Cleveland et al. 1998) is primarily
sensitive to $^8$B neutrinos. The current accuracy of the
experimental measurement of the $^8$B solar neutrino flux is about
9\% (Ahmed et al. 2004), which is much less than the theoretical
uncertainty. Moreover, the uncertainty in the predicted $^8$B
neutrino flux is an important parameter in many analyses of
neutrino oscillation characteristics that make use of solar
neutrino measurements.

 For the p-p, pep, hep, and $^7$Be solar neutrino
fluxes, the difference in the total neutrino uncertainties between
the conservative error estimates (column [2] of
Table~\ref{tab:abundanceuncertainties}) and  the more optimistic
error estimates (column [3] of
Table~\ref{tab:abundanceuncertainties}) are not large enough to
affect the interpretation of planned or ongoing solar neutrino
experiments. Even the total $Z/X$ method gives estimates for the
total flux uncertainties that are similar to the results obtained
by conservative or optimistic assumptions about the individual
abundance uncertainties.

The most abundant neutrino fluxes from the CNO reactions, the
$^{13}$N and  $^{15}$O neutrino fluxes, are very roughly
proportional to the assumed CNO abundances. Since the recently
published  abundance analyses of the volatile elements (including
C, N, and O) are much lower than previously believed, but with a
current quoted error that is relatively small, the total
uncertainties for the CNO neutrino fluxes are about a factor of
two larger when computed using the total $Z/X$ method than the
uncertainties that are obtained using the recent analysis of
abundance uncertainties by Asplund et al. (2005).  Unfortunately,
no precision experiments to measure the CNO solar neutrinos are
currently planned.

If we had combined the O, Ne, and Ar abundance uncertainties
incoherently rather than coherently as was done in constructing
Table~\ref{tab:totaluncertainties} (see discussion in
\S~\ref{subsec:correlation}), the total conservative uncertainty
in the $^8$B neutrino flux would have been 14.4\% instead of our
preferred value of 16.3\%. The differences between the coherent
and incoherent combination of the O, Ne, and Ar abundance
uncertainties are much less important for all the other solar
neutrino fluxes (less than 10\% of the total estimated
uncertainty, a change which is beyond the anticipated experimental
precision).

\section{SUMMARY AND DISCUSSION}
\label{sec:summary}

In this paper, we have computed for the first time the sensitivity
of each solar neutrino flux to the  abundance of each chemical
element on the surface of the Sun. With these computations, we are
able to identify which element abundances most strongly affect the
solar neutrino fluxes.  Moreover, we are able to calculate a
better-founded estimate of the total uncertainty in each neutrino
flux due to composition uncertainties. In previous treatments, all
the uncertainties in the solar chemical composition have been
lumped into a single parameter, the heavy element to hydrogen
ratio, $Z/X$.

The results presented here are important for solar neutrino
research since the flux that is most affected by our more detailed
error treatment, the $^8$B solar neutrino flux, is also the flux
that is measured directly by the SNO and the Super-Kamiokande
solar neutrino experiments.

We summarize in \S~\ref{subsec:partialssummary} our results for
the partial derivatives of each neutrino flux with respect to each
element abundance. We also remark in this subsection on the
historical robustness of the partial derivatives $\partial \ln
\phi/\partial \ln (Z/X)$. We then describe in
\S~\ref{subsec:uncertaintiessummary} our conservative and
optimistic estimates for uncertainties in the individual element
abundances. We summarize our principal results in
\S~\ref{subsec:resultssummary}. We first discuss which individual
element abundances contribute most to the neutrino flux
uncertainties (\S~\ref{subsubsec:individualabundances}) and then
we describe the results when the uncertainties from all elements
are combined (\S~\ref{subsubsec:combined}).  Finally, we summarize
the uncertainties from all known sources of error
(\S~\ref{subsubsec:totalallsourcessummary}).  We present in
\S~\ref{subsec:final} our final word on the subject of solar
abundances and neutrino fluxes.

\subsection{Partial derivatives}
\label{subsec:partialssummary}

The principal calculational tool in our analysis is the set of
partial derivatives, $\alpha_{ij}$, of each solar neutrino flux
with respect to each element abundance. These partial derivatives
are defined in equation~(\ref{eq:definitionalphaij}) and
equation~(\ref{eq:definitionbeta}) and are presented in
Table~\ref{tab:alpha_ij} and Table~\ref{tab:alpha_ij_bp04p}.

 Throughout
this paper, we present  estimates for neutrino flux uncertainties
that use the partial derivatives obtained with the aid of two
different solar models, BP04  (which incorporates the older
Grevesse \& Sauval 1998 solar abundances) and BP04+ (which
incorporates  recent abundance determinations for the volatile
elements, see Table~1 of Bahcall et al. 2004b). The tabulated results in this paper show that the estimated
uncertainties in the neutrino fluxes are essentially the same
whether the partial derivatives $\alpha_{ij}$ are calculated using
the solar model BP04 or the solar model BP04+. In the tables in
the main text, we present without parentheses (with parentheses)
the results obtained using the BP04 solar model (the BP04+) solar
model.

 For comparison with earlier results and for testing the robustness
of  partial derivative calculations over more than two decades, we
present in Table~\ref{tab:alpha_iZoverX} published values from
1982, 1988, and 2004 for the partial derivatives with respect to
$Z/X$. The rms fractional change in the partial derivatives
$\partial \ln \phi/\partial \ln (Z/X)$ is only 2\% when the
partial derivatives of Bahcall \& Ulrich (1988) are compared with
those obtained using the recent solar model BP04. Despite all the
improvements of the solar model since 1988, the partial
derivatives are practically unchanged.

Table~\ref{tab:totalcompositionuncertainties} and
Table~\ref{tab:totaluncertainties} compare in their last columns
the uncertainties computed with the 1988 partial derivatives
$\partial \ln \phi/\partial \ln (Z/X)$ and with the 2004 partial
derivatives.  The total uncertainties due to abundances and the
total uncertainties due to all sources are practically the same
when computed with the old and new derivatives.

\subsection{Abundance uncertainties: conservative and optimistic}
\label{subsec:uncertaintiessummary}

  We estimate conservatively the uncertainty in the element
abundances by comparing the previously standard (Grevesse \&
Sauval 1998)  and the most recent abundance determinations
(Asplund et al. 2005).  We adopt  as $1\sigma$ the differences
between the previous and the recent abundance determinations; this
conservative estimate is our preferred assessment of the abundance
uncertainties. Asplund et al. (2005) give, based upon recent
analyses,   abundance uncertainties that are smaller than our
conservative estimates. We adopt as our optimistic estimate of the
abundance uncertainties the results given by Asplund et al.
(2005).

Table~\ref{tab:abundanceuncertainties} summarizes the adopted
conservative and optimistic uncertainties for the individual
element abundances.

We perform all of our calculations with both the conservative and
with the optimistic estimates for the abundance uncertainties and
compare the results obtained with both sets of abundance
uncertainties.

\subsection{Results}
\label{subsec:resultssummary}

\subsubsection{Flux uncertainties due to individual element
abundances} \label{subsubsec:individualabundances}

We present in Table~\ref{tab:individualuncertainties} and
Table~\ref{tab:optimisticuncertainties} the uncertainty in each
neutrino flux due to uncertainties in the determination of each of
the most important chemical elements. The largest  uncertainty for
both the important $^7$Be and $^8$B neutrino fluxes is due to the
iron abundance, which is strongly ionized in the solar interior
where the neutrinos are produced.  Oxygen, neon, silicon, and
sulphur all contribute significantly to the $^7$Be and $^8$B
neutrino flux uncertainties.

The neon abundance is particularly problematic since it cannot be
measured directly in the solar photosphere and escapes from
meteorites. The abundance uncertainties estimated for neon may be
regarded as best-guesses based upon our limited understanding of
the environments in which neon is detectable in the Sun.

 For the $^{13}$N and $^{15}$O solar neutrino fluxes, the carbon
abundance dominates the composition uncertainty. Carbon is also
the largest contributor to the calculated composition uncertainty
of the basic p-p solar neutrino flux if we adopt the conservative
abundance uncertainties. For the more optimistic abundance
uncertainties of Asplund et al. (2005), iron contributes somewhat
more than carbon to the estimated uncertainty in the p-p neutrino
flux.

\subsubsection{Combined flux uncertainties from all elements}
\label{subsubsec:combined}

Table~\ref{tab:totalcompositionuncertainties} gives for each solar
neutrino flux the total uncertainty from all the abundance
determinations. We have used
equation~(\ref{eq:totalfractionalfluxuncertainty}) to combine the
uncertainties from different chemical elements except for oxygen,
neon, and argon. We combine coherently the abundance uncertainties
from oxygen, neon, and argon (see discussion in
\S~\ref{subsec:correlation}), since the abundances of neon and
argon are measured with respect to oxygen. Our preferred
(conservative) estimated uncertainties are given in the second
column of Table~\ref{tab:totalcompositionuncertainties}, whereas
the more optimistic uncertainties  are presented in column (3).
The conservative uncertainties are, in all cases, more than a
factor of two larger than the optimistic uncertainties.

 We present in the fourth column of
Table~\ref{tab:totalcompositionuncertainties} the larger
uncertainties estimated using the traditional lumping-together of
all abundance uncertainties into a single $\Delta (Z/X)/(Z/X)$.
The $Z/X$ uncertainties are typically somewhat more than a factor
of three larger than the optimistic uncertainties.

The uncertainty for the $^8$B neutrino flux provides the most
dramatic and also the most important example of the differences
between the various ways of calculating the flux uncertainties.
The traditional $Z/X$ method gives a 20\% flux uncertainty for
$^8$B neutrinos. Our conservative estimate using individual
element abundances is 11.6\%. The optimistic estimate for the
$^8$B composition uncertainty is only 5\% using Asplund et al.
(2005) abundance uncertainties.  Thus there is a factor of four
difference in the size of the estimated flux uncertainty depending
upon which method is used to estimate the composition uncertainty.

The uncertainty for the important $^7$Be neutrino flux also varies
by a factor of four depending upon the method of calculation: 8\%
($Z/X$ estimate), 5.3\% (conservative individual abundance
uncertainties), and 2.2\% (optimistic individual abundance
uncertainties).

\subsubsection{The role of correlations}
\label{subsubsec:roleofcorrelations}

Correlations among the uncertainties of the different element
abundances can affect the calculated uncertainties in the solar
neutrino fluxes. The correlations can be taken into account using
equation~(\ref{eq:totalfractionalfluxuncertainty}),
equation~(\ref{eq:radiochemicalfractionaluncertainty}), and
equation~(\ref{eq:meteoriteatmosphere}). We have made crude
estimates of the correlations in the present paper. We have used
$\rho(k,l) = 1$, when $k$ and $l$ are two different elements from
among O, Ne, and Ar. We have neglected all other correlations.
With these estimates, correlations are not numerically important
in determining the final uncertainties in the solar neutrino
fluxes, especially since we advocate using the large, conservative
abundance uncertainties given in the second column of
Table~\ref{tab:abundanceuncertainties}.

It would be  useful  in future compilations of element abundances
to specify explicitly  the correlations between different element
abundances and their uncertainties. When the present conflict
between recent abundance determinations and helioseismology
measurements is resolved and we can adopt more optimistic
estimates for the abundance uncertainties, it will be desirable to
evaluate accurately, using new data, the effects of correlations
on the calculated neutrino fluxes.

\subsubsection{Total neutrino flux uncertainties from all known sources}
\label{subsubsec:totalallsourcessummary}

Table~\ref{tab:totaluncertainties} gives the total uncertainties
from all known sources for each solar neutrino flux and for the
predicted rate of each radiochemical experiment. The uncertainties
for everything except the surface chemical composition of the Sun
are taken from Bahcall \& Pinsonneault (2004).

\subsubsubsection{${\rm ^8B}$ neutrino flux uncertainty}
\label{subsubsubsec:b8}

The uncertainty for the crucial $^8$B neutrino flux is most
affected by the detailed estimate using individual abundances. In
the traditional $Z/X$ method, the total uncertainty in the
calculated $^8$B neutrino flux is a whopping 23.0\% (actually
24.8\% if we use BP04 partial derivatives rather than the Bahcall
\& Ulrich 1988 partial derivatives).  Our preferred conservative
uncertainty estimate is 16.3\%; our optimistic error estimate is
only 12.5\%.

With our preferred conservative estimate, the composition
uncertainty still remains the largest contributor to the
uncertainties in the calculation of the $^8$B neutrino flux (see
Table~2 of Bahcall \& Pinsonneault 2004 for estimates of neutrino
flux uncertainties due to all sources).  The conservative
composition error for the $^8$B neutrino flux is 11.6\% (see
Table~\ref{tab:totalcompositionuncertainties} of this paper),
which is much larger than the next largest contribution to the
uncertainty, 7.5\%, from the low energy cross section factor for
the \hbox{$^3$He($\alpha$, $\gamma$)$^7$Be} nuclear fusion
reaction (see Table~2 of Bahcall \& Pinsonneault 2004).

In order to reduce the composition uncertainty to a level where it
is no longer the largest contributor to the $^8$B neutrino flux
uncertainty, we would need to have confidence in the optimistic
abundance uncertainties. For the optimistic uncertainties, the
composition error for the $^8$B neutrino flux is only 5\%
(Table~\ref{tab:totalcompositionuncertainties}). But, it is
difficult to be confident in the optimistic abundance
uncertainties since, as described in the introduction of this
paper, the new abundance analyses lead to solar models in conflict
with helioseismological measurements.

The reduced estimated uncertainty for the $^8$B solar neutrino
flux  has  implications for solar neutrino research since the
$^8$B neutrino flux is measured by the SNO and Super-Kamiokande
solar neutrino experiments.

The total $^8$B neutrino flux measured by the neutral current mode
of the SNO experiment (Ahmed et al. 2004) is
\begin{equation}
\phi(^8{\rm B, SNO}) ~=~ 0.90 \phi(^8{\rm B,
BP04~solar~model})\left[ 1.0 \pm 0.09 \pm 0.16 \right] \, ,
\label{eq:b8snovsstandardmodel}
\end{equation}
where the first uncertainty listed in
equation~(\ref{eq:b8snovsstandardmodel}) is the $1\sigma$
measurement error and the second (larger) uncertainty is the
estimated $1\sigma$ uncertainty in the solar model calculation
(taken from Table~\ref{tab:totaluncertainties}).  If all the data
from solar neutrino and reactor experiments are combined together,
the above relation becomes (Bahcall et al. 2004a):\nocite{BGP04}
\begin{equation}
 \phi(^8{\rm B, SNO}) ~=~ 0.87 \phi(^8{\rm B,
 BP04~solar~model})\left[ 1.0 \pm 0.05 \pm 0.16 \right] \, .
 \label{eq:b8fitvsstandardmodel}
 \end{equation}

The calculated $^8$B neutrino flux (Bahcall \& Pinsonneault 2004)
agrees with the measured flux to better than $1\sigma$.  The
theoretical uncertainty is much larger than the uncertainty in the
measurements.

\subsubsubsection{${\rm ^7Be}$ neutrino flux uncertainty}
\label{subsubsubsec:be7}

 For the $^7$Be solar neutrinos, which will be measured by the
BOREXINO solar neutrino experiment (see Alimonti et
al. 2002)\nocite{borexino},
the situation is somewhat different.  The largest contribution to
the presently estimated uncertainty in the predicted flux is 8.0\%
from the laboratory measurement of the rate of the
\hbox{$^3$He($\alpha$, $\gamma$)$^7$Be} nuclear fusion reaction
(see Adelberger, et al. 1998)\nocite{adelberger}. Even our
conservative composition uncertainty is only 5.3\% for the $^7$Be
neutrino flux (Table~\ref{tab:totalcompositionuncertainties}).
However, the situation may change in the near future. A recent
measurement of the \hbox{$^3$He($\alpha$, $\gamma$)$^7$Be}
reaction by Singh et al. (2005)\nocite{singh} reports a precision for this
reaction rate which is much better than 5\%.  If the Singh et al. (2005)
result is confirmed by future measurements, then even for the
$^7$Be solar neutrinos the solar composition will be the largest
contributor to the calculational uncertainty.

\subsubsubsection{{\rm p-p, pep},  ${\rm ^{13}N}$, and ${\rm ^{15}O}$
  neutrino flux
uncertainties} \label{subsubsubsec:otherfluxes}

The conservative composition uncertainty is the largest
contributor to the estimated uncertainty in the calculation of the
p-p, pep, $^{13}$N, and $^{15}$O solar neutrino fluxes (compare
Table~\ref{tab:totalcompositionuncertainties} of this paper with
Table~2 of Bahcall \& Pinsonneault 2004). However, the optimistic
composition uncertainties are less than other uncertainties in
calculating these neutrino fluxes.

The p-p solar neutrino flux has been measured by combining the
results from all the relevant solar neutrino and reactor
experiments, together with the imposition of the luminosity
constraint (Bahcall 2002)\nocite{luminosity}.  The result is (Bahcall
et al. 2004a)\nocite{BGP04}
\begin{equation}
\phi({\rm \hbox{p-p}, all~neutrino~experiments}) ~=~ 1.01 \phi({\rm \hbox{p-p},
BP04~solar~model})\left[ 1.0 \pm 0.02 \pm 0.01 \right] \, ,
\label{eq:ppexperimenttheory}
\end{equation}
where the first uncertainty listed in
equation~(\ref{eq:ppexperimenttheory}) is the $1\sigma$
measurement error and the second (smaller) uncertainty is the
estimated $1\sigma$ uncertainty in the solar model calculation
(taken from Table~\ref{tab:totaluncertainties}).

\subsubsubsection{Comparisons of Total Uncertainties from
Different Methods} \label{subsubsubsec:totaluncertainties}

Table~\ref{tab:totaluncertainties} shows that for all the solar
neutrino fluxes produced in the p-p chain except the $^8$B
neutrino flux (i.e., the p-p, pep, hep, and $^7$Be neutrino
fluxes), the total uncertainties from all sources are similar for
the conservative and the $Z/X$ estimates for the composition
uncertainties. Of course, this must be the case if the composition
uncertainties are only a small contribution to the total
uncertainties. Indeed for hep neutrinos, the calculated rate of
the fusion reaction is the dominant recognized uncertainty, 15.1\%
(see Bahcall and Pinsonneault 2004; Park et al. 2003\nocite{park03}).  But, as we
have seen above, the composition uncertainties are at present the
largest contributor to the total uncertainties for all neutrino
fluxes except the hep and the  $^7$Be neutrinos.

The reason that the conservative and the $Z/X$ methods give such
similar results is that the recent change in the abundances of the
volatile elements affects in a similar way the historical average
of $Z/X$ and the difference between previous standard and current
element abundances.  The recently inferred changes in the volatile
element abundances represent the largest modern revision in the
solar element abundances from one epoch of abundance
determinations to its successor.

\subsection{Do neutrino measurements tell us which abundances are
correct?}
\label{subsec:whichcorrect}

The values of the solar neutrino fluxes predicted by solar models
depend upon the assumed heavy element abundances. Unfortunately,
the agreement between solar neutrino predictions and solar
neutrino measurements is excellent for calculations made with both
the Grevesse and Sauval (1998) abundances and with the Asplund et
al (2005) abundances. In both cases, the differences between
measurements and predictions are smaller than the uncertainties in
the predictions and measurements (Bahcall \& Pinsonneault 2004;
Bahcall,  \& Basu Serenelli 2005\nocite{BS05}).

In fact, Table~1 of Bahcall \& Pinsonneault (2004) gives the
neutrino fluxes for a  state-of-the art solar model that was
computed using the older Grevesse \& Sauval (1998) abundances
(BP04) and the more recent Asplund et al. (2005) abundances.  If
we compare the fluxes from these two models, we see that in all
cases their neutrino fluxes agree to within $1\sigma$ of the total
theoretical uncertainties given in the first column of
Table~\ref{tab:totaluncertainties}.

\subsection{Final word}
\label{subsec:final}

We conclude, as we began, by emphasizing that composition
uncertainties are the most important uncertainties for determining
the accuracy of solar neutrino calculations. It would be desirable
if multiple groups would undertake determinations of solar
abundances in order that systematic uncertainties can be more
readily assessed. The possibility of obtaining improved solar
spectroscopic data should also be considered. In the future, the
correlations between the uncertainties in different element
abundances should be specified explicitly by researchers
determining the abundances. The correlations, which hopefully will
become relevant for precise evaluations of reduced neutrino flux
uncertainties, can be taken into account using the formulae given
in this paper.

\acknowledgments J. N. B. and A. M. S. are supported in part by NSF grant
PHY-0070928.  We
are grateful to B. Draine, N. Grevesse, K. Lodders, C. Pena-Garay,
M. Pinsonneault, and D. Sasselov for valuable discussions. We are
grateful to E. Lisi and M. Pinsonneault for insightful comments on
a draft of the manuscript.

\end{document}